# (Electro)catalytic oxidation of sulfide and recovery of elemental sulfur from sulfide-laden streams

*Natalia Sergienko[a,b], Elizabeth Cuervo Lumbaque[a,b], Jelena Radjenovic [a,c*]*

[a]*Catalan Institute for Water Research (ICRA-CERCA), Emili Grahit 101, 17003 Girona, Spain*

[b] *University of Girona, Girona, Spain*

[c]*Catalan Institution for Research and Advanced Studies (ICREA), Passeig Lluís Companys 23, 08010 Barcelona, Spain*

*\* Corresponding author:*

*Jelena Radjenovic, Catalan Institute for Water Research (ICRA), c/Emili Grahit, 101, 17003 Girona, Spain*

Phone: + 34 972 18 33 80; Fax: +34 972 18 32 48; E-mail: jradjenovic@icra.cat

## Abstract

$Mn_xO_y$ coated over $TiO_2$ nanotube array substrate was doped with Mo and polyaniline (PANI) and applied for electrochemical desulfurization of concentrated sulfide ($HS^-$) solutions at basic pH, typical of biogas scrubbing solutions and industrial wastewater. Mo and PANI co-dopants significantly enhanced the anode activity towards sulfide oxidation and ensured its complete stability even in highly corrosive sulfide solutions (e.g., 200 mM $HS^-$). This was due to the increased electrochemically active surface area, improved coating conductivity and reduced charge transfer resistance. The (electro)catalytic oxidation of $HS^-$ demonstrated robust performance with very limited impact of different operational parameters (e.g., dissolved oxygen, anode potential, $HS^-$ concentration). Due to the formation of elemental sulfur ($S^0$) layer at the anode surface at basic pH, longer term anode usage requires its periodic removal. Chemical dissolution of $S^0$ with toluene allows its rapid removal without affecting the anode activity, and easy recrystallization and recovery of pure sulfur.

**Keywords**: Manganese oxide, molybdenum doping, polyaniline, electrocatalysis, sulfide oxidation

## 1. Introduction

Over the last few decades, extensive anthropogenic activity led to a drastic increase in atmospheric sulfur [1]. Elevated content of sulfur in the atmosphere deteriorates air quality, causes acid rains, and influences aerosol radiative forcing [2]. Considering the severity of these environmental concerns, anthropogenic sulfur emission must be controlled and minimized. Various industrial processes, including oil refining, mining, tanning, wood pulp processing and food processing contribute to sulfur emissions [3-5]. These industrial activities generate waste streams that contain high concentrations of sulfide, which can volatilize from wastewater into the atmosphere [6]. Besides that, fuels used for energy generation (e.g., fossil or biogas) often contain substantial amounts of sulfide [7]. Upon combustion, sulfide is oxidized and released into the atmosphere as sulfur dioxide ($SO_2$). Combustion of the non-desulfurized fuels is the dominant contribution to the total anthropogenic sulfur emissions [8]. Hence, excessive presence of sulfur in the atmosphere can be greatly diminished by ensuring complete desulfurization of fuels prior to combustion, as well as sulfide removal from the industrial waste streams.

Biogas is a fuel where hydrogen sulfide concentration can reach particularly high levels [9]. Sulfate reducing bacteria (SRB), a member of the microbial consortia in anaerobic digestors, actively utilize sulfate in the feed as electron acceptor [10]. As part of their anaerobic respiration process, SRB produce hydrogen sulfide, which can volatilize from the liquid phase and mix with methane [11]. Depending on the sulfate load and pH of the reactor, hydrogen sulfide concentration in raw biogas may vary from 50 to 5000 ppm [12]. In addition to serious environmental implications associated with the generation of $SO_2$ upon combustion, high hydrogen sulfide content (i.e., above 500 ppm) causes severe corrosion of piping, pumps, gas storage tanks and combustion engines [12, 13]. The conversion of biogas into biomethane suitable for storage, distribution within the natural gas grid, or combustion without sulfur dioxide generation requires upgrading the biogas to contain 95-99% $CH_4$ and 1-5% $CO_2$, while completely removing any traces of $H_2S$. [14].

A variety of technologies was developed to minimize hydrogen sulfide presence in the gas stream [15, 16]. Although physicochemical desulfurization (e.g., Claus process, adsorption, absorption in aqueous solutions, chemical scrubbing, membrane separation) is a well-established approach that demonstrates high efficiency in hydrogen sulfide removal from the natural gas, it is characterized by the high chemical and energy consumption [16]. Furthermore, most of these methods require gas stream with high hydrogen sulfide content at high flow through velocity, which limits their application to the natural gas treatment [17]. Biological desulfurization is a more appropriate technology for the biogas treatment as it is effective for the low- and medium-strength sulfide gas streams [18]. Biological removal of hydrogen sulfide is typically achieved in two steps: *i)* absorption of $H_2S$ from the biogas in alkaline solution and its dissolution to bisulfide ($HS^-$), and *ii)* oxidation of $HS^-$ to elemental sulfur, $S_0$, by the sulfate oxidizing bacteria (SOB) [19]. Despite the excellent sulfide removal efficiency and sulfur recovery achieved with biological desulfurization, this technology still has two significant limitations. First, extremely low sulfide content (i.e., <0.2 mM) leads to the process failure as SOB are unable to tolerate it. Second, biological sulfide oxidation to elemental sulfur is not 100% selective, leading to a gradual accumulation of the dissolved sulfur species (i.e., $S_2O_3^{2-}$, $SO_4^{2-}$), which ultimately also leads to the process failure [20]. Finally, biological desulfurization needs continuous addition of caustic, nutrients and air sparging, which drastically increases the operating costs [21].

In our previous study, we demonstrated that electrocatalytic sulfide oxidation based on manganese oxide-coated Ti plate with $TiO_2$ nanotube array (NTA) interlayer is capable of selective and complete oxidation of $HS^-$ to $S_0$ in municipal sewage [22]. The excellent performance in terms of efficiency and selectivity is due to the high affinity of manganese oxides towards sulfide oxidation to elemental sulfur. Application of a very low anodic potential of (e.g., 0.6 V/SHE) enabled the re-oxidation of the reduced $Mn_xO_y$, thus ensuring that the excellent catalyst activity is maintained. The neutral pH, typical of municipal sewage, enabled the desorption of elemental sulfur particles, as opposed to a layer of elemental sulfur formed at the anode surface at more basic pH values (i.e., pH 12). In addition to that, electrocatalytic desulfurization can be achieved without the addition of chemicals and can effectively oxidize sulfide even at very low concentrations, thus overcoming major limitations of the biological processes for the removal of hydrogen sulfide. Considering all the advantages of the proposed electrochemical approach, its excellent performance can be expanded beyond wastewater desulfurization to the treatment of various waste streams, including sulfide-laden industrial effluents and biogas scrubbing solutions. However, prior to the application of the electrocatalytic sulfide oxidation based on Ti/TiO$_2$ NTA Mn$_x$O$_y$ to industrial sulfide-laden wastewater or scrubbing solutions, several limitations of this approach must be addressed. First, high concentration of sulfide in these streams, in the order of several hundred mg L$^{-1}$, leads to the Mn$_x$O$_y$ dissolution because the regeneration of the catalyst under anodic polarization is not fast enough. To increase the stability of the catalyst at high sulfide concentration, the coating was modified by the incorporation of Mo and polyaniline (PANI) into the Mn$_x$O$_y$ host lattice. As we demonstrated in our recent study, cationic doping of Mn$_x$O$_y$ with Mo enhances the catalytic activity, stability, and electrical characteristics (i.e., conductivity and electroactive surface area) of the anode through the formation of the redox couples between Mn and Mo (i.e., Mn III, Mn IV and Mo IV, Mo VI), and generation of oxygen vacancies in the Mn$_x$O$_y$ lattice [23-25]. PANI is a conductive polymer and its incorporation into the Mn$_x$O$_y$ is expected to boost the conductivity of the coating and further enhance the electrode´s stability [26-28]. Second, both sulfide-laden industrial wastewater and biogas scrubbing solutions are characterized by the high pH (i.e., >pH 10) [19, 29]. Basic pH will prevent the desorption of the sulfur particles from the surface, which will complicate sulfur recovery and lead to anode passivation. To address this issue, we tested different physico-chemical sulfur methods to extract sulfur from the used anode and thus restore the catalytic activity of the $Mn_xO_y$ coating.

## 2. Materials and methods

### 2.1. The synthesis of TiO2 NTA $Mn_xO_y$-Mo-PANI anodes

5 × 4 cm Ti mesh (not coated, 99.6% purity, De Nora, Italy) underwent sandblasting and was then degreased by sonication in isopropanol, acetone, and methanol. It was subsequently rinsed with deionized water. After drying the mesh with a gentle flow of nitrogen, it was subjected to etching in a 17% w/w hydrochloric acid (HCl) aqueous solution (Scharlab, Spain) at 75°C for 15 min. Next, the mesh was anodized in a mixture of glycerol and deionized water with a volume ratio of 50:50, which also contained 0.5 wt.% ammonium fluoride ($NH_4F$) (Scharlab, Spain). To carry out the anodization, the titanium mesh was polarized against a split stainless steel mesh. The anodization process began by gradually increasing the potential from the open circuit (OC) to 20 V, which was then maintained constant for a period of 2 hours. The potential of the electrochemical cell was controlled using an Autolab 302N potentiostat/galvanostat equipped with a voltage amplifier (Metrohm Autolab B.V., The Netherlands). After anodization, the resulting NTA samples were soaked in ethanol and subjected to calcination at 400°C under an argon atmosphere for 2 hours in a tubular oven (Nabertherm, Germany).

The Ti/$TiO_2$ NTA was then coated through anodic electrodeposition. The procedure was carried out in a three-electrode setup at an ambient temperature of 24±1°C. A stainless steel mesh served as the counter electrode, while an Ag/AgCl electrode (KCl 3M, Bioanalytical systems, the Netherlands) acted as the reference electrode. The electrodeposition medium contained 0.1 M MnSO$_4$ and 0.5 M H$_2$SO$_4$. Depending on the synthesized material, the electrolyte was also amended with 100 µM Na$_2$MoO$_4$, or 100 µM Na$_2$MoO$_4$ and 100 µM aniline for the synthesis of TiO$_2$ NTA Mn$_x$O$_y$-Mo or TiO$_2$ NTA Mn$_x$O$_y$-Mo-PANI, respectively. The electrodeposition process was conducted using the potentiostatic mode, maintaining a potential of 1.7 V/SHE (vs Standard Hydrogen Electrode). All potentials mentioned in this work are expressed relative to the Standard Hydrogen Electrode (SHE) and were calculated using the Nernst equation. The loading of the coating was calculated based on Faraday's law and the reproducibility of each deposition was ensured by limiting applied charge to 13 C.

## 2.2. Material characterization

The surface morphology of the synthesized materials was studied using an ultra-high-resolution field emission scanning electron microscopy (FESEM) (The Magellan 400L, FEI, US). The chemical state analysis of the Mn$_x$O$_y$ coating was performed using X-ray photoelectron spectroscopy (XPS) (PHOIBOS 150, Specs, Germany).

The estimation of the electroactive surface area of the synthesized electrodes involved measuring the double layer capacitance (C$_{dl}$) observed during cyclic voltammetry (CV) measurements in a 100 mM NaNO$_3$ solution. Voltametric scans were conducted within a potential window of 0.4 to 1 V/SHE, at scan rates ranging from 5 to 1 mV s$^{-1}$. The C$_{dl}$ values were determined through the analysis of linear regression of the current versus the scan rate, following the equation:

$$\frac{I_a - I_c}{2} = C_{dl}\nu \qquad \text{(eq. 1)}$$

where I$_a$ and I$_c$ are the anodic and cathodic currents observed in the forward and reversed scans (mA), respectively, and ν denotes the applied scan rate (mV s$^{-1}$). The electroactive surface area was then determined by dividing the measured capacitance by the areal capacitance of a planar metal oxide based electrode (i.e., 60 mF cm$^{-2}$) [30].

Electrochemical impedance spectroscopy (EIS) of the synthesized material electrodes was carried out in 100 mM NaNO$_3$. A potential of 1.4 V/SHE was applied within a frequency range of 0.1 Hz to 50 kHz. The results of these measurements were plotted in a Nyquist plot.

The crystal structure of sulfur particles was studied with an X-ray powder diffraction (X'Pert MPD, PANalytical, Netherlands).

## 2.3. Sulfide removal tests

Electrochemical experiments were performed in a non-divided glass cell (100 mL) with an air-tight seal. The synthesized materials were employed as anodes, while stainless steel mesh was used as a counter (Advent Research Materials, UK). The anode was polarized versus Ag/AgCl (3M KCl) reference electrode (Advent Research Materials, UK). and Ag/AgCl (3M KCl) and stainless steel mesh (Advent Research Materials, UK) served as the reference and counter electrodes, respectively. Considering that industrial wastewater and scrubbing solutions are characterized by the high ionic strength and basic pH, sulfide removal experiments were performed in 100 mM NaCO$_3$ and 20 mM HS$^-$ at pH ∼ 12 unless stated otherwise. The anode was polarized at 1.4 V/SHE and experiments were performed in oxygen free environment, unless stated otherwise. The absence of oxygen in the system was achieved by purging the electrolyte with the nitrogen gas prior to the experiment and maintaining a gentle nitrogen purge of the reactor headspace during the experiment.

First, the effect of the dopants (i.e., Mo or PANI) on the sulfide removal kinetics was evaluated in the electrocatalytic experiments using Ti/TiO$_2$ NTA Mn$_x$O$_y$, Ti/TiO$_2$ NTA Mn$_x$O$_y$-Mo and Ti/TiO$_2$ NTA Mn$_x$O$_y$-Mo-PANI anodes. To compare the performance of these materials with the conventional mixed metal oxide (MMO) electrode, additional experiments were carried out using

Ti/IrO$_2$-Pt mesh anode (Magneto Special Anodes, NL) in the same experimental set up. For the best-performing Ti/TiO$_2$ NTA Mn$_x$O$_y$-Mo-PANI anode, we investigated the impact of several operational parameters on the electrocatalytic sulfide removal and final products of its electrooxidation. First, the role of oxygen was elucidated by comparing the sulfide oxidation kinetics in the oxygenated and deoxygenated supporting electrolytes. Sulfide oxidation rate by oxygen was determined by gently purging 20 mM HS$^-$ and 100 mM NaCO$_3$ solution with oxygen in the absence of the anode. Second, the influence of different potentials on the removal of sulfide was examined by conducting experiments under open circuit (OC) conditions, as well as at at 1.4 V, 1.6 V or 1.8 V/SHE. The effect of the supporting electrolyte was evaluated by conducting the experiments in Na$_3$PO$_4$, NaOH, Na$_2$CO$_3$ and NaCl supporting electrolytes at the same ionic strength (i.e., 100 mM) and pH∼12. To gain insight into the influence of the ionic strength, electrocatalytic sulfide removal was investigated in 20 mM, 100 mM, and 500 mM Na$_2$CO$_3$ supporting electrolytes. Lastly, we investigated the impact of varying initial sulfide concentrations (i.e., 2 mM, 20 mM and 200 mM) on its electrooxidation kinetics, to ensure the applicability of the process in a wide range of HS$^-$ concentrations.

The energy consumption, expressed as electric energy per order (W h L$^{-1}$), quantifies the amount of energy required to decrease the sulfide concentration by one order of magnitude within a specific volume of the treated solution, as described by the following equation:

$$E_{EO} = \frac{V \cdot I \cdot t}{v \cdot \log\left(\frac{C_0}{C}\right)} \quad \text{(eq. 2)}$$

where V is cell voltage (V), I is measured current (A), v is the cell volume (L), C$_0$ and C are the initial and final concentration of sulfide, and t is the electrolysis time (h).

The effect of sulfur accumulation at the Ti/TiO$_2$ NTA Mn$_x$O$_y$-Mo-PANI anode on sulfide oxidation kinetics and final product distribution was evaluated in three consecutive applications of the same anode. Several sulfur extraction strategies such as physical cleaning by sonication, and chemical cleaning with ethanol or toluene in between cycles were evaluated.

## 2.4. Sample analysis

Inductively coupled plasma-optical emission spectrometry (ICP-OES) was employed to monitor the concentration of total dissolved manganese in the supporting electrolyte (Agilent 5100, Agilent Technologies, US). Ion chromatography (IC, Dionex IC5000 (Dionex, USA) was employed to determine the concentration of dissolved sulfur species, including $HS^-$, $S_2O_3^{2-}$ and $SO_4^{2-}$. At pH>9, the majority of sulfite ($SO_3^{2-}$) is converted to sulfate (i.e., $SO_4^{2-}$), therefore the presence of $SO_3^{2-}$ was discarded. To determine the concentration of polysulfide (polyS), all dissolved sulfur species were oxidized to sulfate with hydrogen peroxide ($H_2O_2$). To ensure a complete oxidation of all sulfur species and avoid the formation of elemental sulfur ($S_0$) because of the sulfide oxidation with $H_2O_2$, the solution was also exposed to UV light for 2 h. The difference between the $SO_4^{2-}$ measured after the UV/$H_2O_2$ oxidation, and the dissolved sulfur species (i.e., $HS^-$, $S_2O_3^{2-}$, $SO_4^{2-}$) measured before the treatment was regarded as polyS. Elemental sulfur was determined as the difference between the total sulfide added and the dissolved sulfur species measured. As no particles could be observed in the solution, all generated elemental sulfur was assumed to be electrodeposited at the electrode surface. All experiments were conducted in duplicates, and the average sulfide concentrations obtained were normalized against the initial values. These normalized values were then fitted to a first-order kinetics model, ensuring a coefficient of determination ($R^2$) of at least 0.99. To compare the sulfide oxidation kinetics observed using different materials (e.g., $Mn_xO_y$, $Mn_xO_y$-Mo and $Mn_xO_y$-Mo-PANI coatings), the reaction rate was normalized to the electroactive surface area.

## 3. Results and discussion

### *3.1. Characterization of the Ti/TiO$_2$ NTA Mn$_x$O$_y$-Mo-PANI electrodes*

In our previous work [22], we provided a detailed description of the interaction between the coating and the Ti/TiO$_2$ NTA, as well as the influence of Mo on the characteristics of Mn$_x$O$_y$. Briefly, the anodization procedure resulted in a uniform and well-aligned layer of TiO$_2$ nanotubes with an outer diameter ranging from 80 to 100 nm, a wall thickness of 7-10 nm, and a length of approximately 1 μm [22]. By employing the electrodeposition procedure, the nanotubes were successfully filled with Mn$_x$O$_y$, enabling a strong interaction between the substrate and the coating. The incorporation of Mo into the Mn$_x$O$_y$ host lattice induced significant structural modifications to the catalyst. First, it promoted the formation of O$_2$ vacancies in the Mn$_x$O$_y$ lattice, thereby enhancing the mobility of lattice oxygen and facilitating the generation of active surface oxygen species. Second, it caused a reduction of the average oxidation state of the Mn$_x$O$_y$ by increasing the presence of Mn III sites, which allowed the formation of redox couples involving Mn and Mo species (i.e., Mn IV/Mn III and Mo IV/Mo VI) [23].

The incorporation of PANI into the Mn$_x$O$_y$ catalyst was first confirmed with the SEM images. As can be seen from **Figure 1**, which depicts pure Mn$_x$O$_y$, Mn$_x$O$_y$-Mo and Mn$_x$O$_y$-Mo-PANI coatings, PANI-doped catalyst is characterized by the clearly distinct morphology. Whereas Mn$_x$O$_y$ and Mn$_x$O$_y$-Mo are comprised of compact micro grains (**Figure 1a, 1b**), Mn$_x$O$_y$-Mo-PANI coating is porous (**Figure 1c**). The porous morphology of the PANI-doped material can be attributed to the mechanism of aniline polymerization and PANI incorporation into Mn$_x$O$_y$ host lattice. Oxidative polymerization of aniline can be achieved either electrochemically or chemically through the addition of an oxidizing agent [31, 32]. Oxidative polymerization of aniline monomer initiates a chain reaction leading to the formation of regular macromolecules, which are then assembled into chain structures [31]. The growth of PANI chains in the acidic media occurs at the potential above 1.05 V/SHE [33]. Considering the electrodeposition potential of 1.5 V/SHE and high oxidizing potential of the $Mn_xO_y$ coating (i.e., 1.23 V/SHE), aniline can polymerize via electrochemical and/or chemical route. In the chemical oxidative polymerization of aniline, $Mn_xO_y$ is reduced into soluble $Mn^{2+}$ and released back into the electrolyte, thus acting as a sacrificial template for PANI structures [34]. Local dissolution of the electrodeposited $Mn_xO_y$ caused by its reduction with the aniline monomer explains the characteristic porous morphology observed in the FESEM image (**Figure 1c**). This suggests that PANI was mainly incorporated by the chemical oxidation with the $Mn_xO_y$, rather than via electropolymerization route. This assumption is further supported by the fact that once the aniline concentration in the electrodeposition bath was increased (i.e., from 100 μM to 1000 μM), the obtained film only contained PANI without any trace of $Mn_xO_y$ in it, implying that the electrodeposited $Mn_xO_y$ was completely dissolved.

In our previous work, we demonstrated that the incorporation of Mo into the $Mn_xO_y$ coating leads to the increase of the electrochemically active surface area of the resulting coating [23]. Based on the measurement of the $C_d$ of the pure and Mo-doped $Mn_xO_y$ anodes, Mo doping led to 1.2-fold increase of the electroactive surface area when compared with the pure $Mn_xO_y$, i.e., 1797 cm$^2$ for pure $Mn_xO_y$ and 2209 cm$^2$ for $Mn_xO_y$-Mo (**Figure S1**). The porous morphology observed for the $Mn_xO_y$-Mo-PANI coating is expected to increase the electrochemically active surface area even further. Indeed, the estimated electrochemically active surface area for the $Mn_xO_y$ sample doped with PANI was 3370 cm$^2$, 1.5 times higher compared with the $Mn_xO_y$-Mo anode (**Figure S1**).

The chemical state of the elements in the $Mn_xO_y$-Mo-PANI coatings was thoroughly examined through XPS analysis (**Figure 2**). Analysis of the Mn 2p spectra revealed two distinct peaks at 653.8 eV and 642.1 eV, exhibiting a spin orbit splitting of approximately 11.7 eV (**Figure 2a**). The asymmetry observed in the Mn $2p_{3/2}$ peak indicates the presence of Mn in mixed valence states, specifically Mn III and Mn IV [35]. The average valence state of Mn was determined based on the value of the Mn 3s spin orbit splitting. The calculated ΔE value of 5.1 eV corresponds to an average oxidation state of 3.3. This determination was based on a linear relationship between the Mn oxidation state and ΔE values, as previously reported in the literature (**Figure S2**) [36]. This average oxidation state is lower compared with the pure electrodeposited $Mn_xO_y$, which is typically characterized by an oxidation state of 4 [37]. However, as can be seen from **Figure S3**, the increased presence of Mn III sites in the $Mn_xO_y$-Mo-PANI coating occurred because of the Mo doping, and not due to the PANI incorporation. Hence, the incorporation of PANI into the $Mn_xO_y$ host lattice did not affect the average oxidation state of the catalyst.

The presence of PANI in the $Mn_xO_y$ was clearly indicated by the C1s and N1s spectra (**Figure 2, Figure S4, Table S1**). The C1s peak of $Mn_xO_y$-Mo-PANI showed a slight shift compared to the $Mn_xO_y$-Mo (from 283.3 eV for $Mn_xO_y$-Mo to 284.5 eV $Mn_xO_y$-Mo-PANI), which can be attributed to the interaction between the $Mn_xO_y$ and PANI [38]. In addition to the shift of the main peak (i.e., C-C/C=C), the incorporation of PANI affected the distribution of the carbon species. The obtained C1s spectra for both $Mn_xO_y$-Mo and $Mn_xO_y$-Mo-PANI exhibited three peaks at 283.3 – 284.5 eV, 284.4 – 285.2 eV and 287.2 – 288.0 eV ascribed to C-C/C=C, C-O/C-N and C=O/C=N components, respectively. Relative concentration of C-O/C-N and C-O/C=N was increased from 17.6% to 47.96%, which clearly indicates the incorporation of PANI into the $Mn_xO_y$. Furthermore, the XPS spectra of the $Mn_xO_y$-Mo-PANI sample demonstrates a well-defined N1s peak, which is absent in the $Mn_xO_y$-Mo sample (**Figure 2, Figure S4**). The N1s peak deconvolution can reveal the redox state of PANI. There are three possible oxidation states for PANI: the reduced state of leucoemeraldine, the half-oxidized state of emeraldine, and the oxidized state of pernigraniline [31].

The first peak observed at 398.7 eV was attributed to the neutral imine (=N-), which is associated with the quinoid group. The second peak at 399.8 eV was ascribed to the neutral amine (-NH-), which is typically bound with the benzenoid group. Finally, the peak with the highest energy was associated with the positively charged nitrogen ($N^+$) [39]. The ratio of the imine to amine nitrogen is often used to determine the redox state of PANI [40, 41]. The relative quantities of =N-, -NH-, $N^+$ were determined to be 30.5%, 64.07% and 5.47%, respectively (**Figure 2, Table S1**), which is characteristic for PANI in the emeraldine state [40]. Since emeraldine is the most conductive form of PANI, its presence in the $Mn_xO_y$ coating is expected to improve the electrical conductivity of the $Mn_xO_y$ [42]. This improved electron transfer achieved by the PANI incorporation was confirmed with the EIS measurements. **Figure S5** depicts the Nyquist diagrams for the Ti/TiO$_2$ NTA $Mn_xO_y$, Ti/TiO$_2$ NTA $Mn_xO_y$-Mo and Ti/TiO$_2$ NTA $Mn_xO_y$-Mo-PANI anodes obtained in 100 mM NaNO$_3$ electrolyte at 1.4 V/SHE. The presence of Mo significantly reduced the charge transfer resistance of the $Mn_xO_y$ coating (i.e., from 180 Ω for pure $Mn_xO_y$ to 12 Ω for Mo doped $Mn_xO_y$), while doping with PANI decreased the resistance even further (i.e., 9 Ω for Mo and PANI doped $Mn_xO_y$ coating) (**Figure S5**).

### *3.2. (Electro)catalytic activity of Ti/TiO$_2$ NTA $Mn_xO_y$ anodes and impact of Mo and PANI doping on sulfide oxidation*

The activity of each synthesized $Mn_xO_y$-based material towards sulfide oxidation was evaluated in chronoamperometric experiments at 1.4 V/SHE in 20 mM HS$^-$ solutions. As can be seen from **Figure 3a,** the non-doped Ti/TiO$_2$ NTA $Mn_xO_y$ anode caused a substantial decrease (i.e., 49.1%) of sulfide concentration within 2 hours of the experiment. Even without any doping, Ti/TiO$_2$ NTA $Mn_xO_y$ anode demonstrated higher activity towards sulfide oxidation when compared to the conventional

Ti/IrO$_2$-Pt anode, which was only capable of 36.2% sulfide removal. The removal rate normalized to the geometrical surface area of the Ti/TiO$_2$ NTA-Mn$_x$O$_y$ was four times faster compared with the Pt-IrO$_x$ anode (i.e., 0.2 m$^3$ h$^{-1}$ m$^{-2}$ for Ti/TiO$_2$ NTA Mn$_x$O$_y$ and 0.051 m$^3$ h$^{-1}$ m$^{-2}$ for Pt-IrO$_x$). Furthermore, as can be seen from **Figure 3b**, Ti/TiO$_2$ NTA Mn$_x$O$_y$ is more selective towards the elemental sulfur compared with the Ti/IrO$_2$-Pt anode. Although the major product for both Ti/TiO$_2$ NTA Mn$_x$O$_y$ and Ti/IrO$_2$-Pt anodes was elemental sulfur, sulfide oxidation on the Ti/IrO$_2$-Pt anode generated more dissolved sulfur species (i.e., 15.4% at Ti/TiO$_2$ NTA Mn$_x$O$_y$ anode vs. 22% at Ti/IrO$_2$-Pt anode, respectively). Better selectivity towards sulfur generation can be attributed to the difference in the sulfide oxidation mechanism on the Ti/TiO$_2$ NTA Mn$_x$O$_y$ and Ti/IrO$_2$-Pt anodes. Sulfide removal on the Ti/IrO$_2$-Pt anode was achieved via electrochemical oxidation [43], whereas the Ti/TiO$_2$ NTA Mn$_x$O$_y$ anode was capable of catalytic oxidation of HS$^-$ due to its interaction with the Mn$_x$O$_y$, which occurs according to the following pathway [44]:

$$Mn^{IV} + HS^- \rightleftharpoons Mn^{IV}S^- + H_2O \quad \text{(eq. 3)}$$

$$Mn^{IV}S^- \rightarrow Mn^{II}S \quad \text{(eq. 4)}$$

$$Mn^{II}S \rightarrow Mn^{II} + S^0 \quad \text{(eq. 5)}$$

The incorporation of Mo into the Mn$_x$O$_y$ coating improved the performance of the system in multiple ways. The presence of Mo increased the selectivity of sulfide oxidation to elemental sulfur, as the yield of the dissolved sulfur species was decreased from 15.4% for pure Mn$_x$O$_y$ to 8.3% for Mn$_x$O$_y$-Mo, and the generation of sulfate was completely suppressed (**Figure 3b**). Furthermore, Mo doping had a positive impact on the sulfide oxidation, improving the removal efficiency from 49.1% to 65%. Higher catalytic activity of the Mo-doped Mn$_x$O$_y$ could arise from the structural changes in the Mn$_x$O$_y$ lattice induced by the Mo incorporation. First, Mo doping increases the electrochemically active surface area of the material (i.e., 1797 cm$^2$ for pure Mn$_x$O$_y$ and 2209 cm$^2$ for Mn$_x$O$_y$-Mo), thus providing more sites available for sulfide oxidation [23, 24]. Second, as a result of the increased number of oxygen vacancies and higher oxygen mobility around the $Mn_xO_y$ lattice caused by the Mo incorporation, and formation of redox couples between Mn and Mo species, the electrical conductivity of the Mo-doped material is higher than that of pure $Mn_xO_y$ [45, 46]. Poor conductivity of the $Mn_xO_y$ limits its reoxidation back to the initial valence state after the reaction between the catalyst and sulfide [38]. Since Mn II can be spontaneously released into the electrolyte as $Mn^{2+}$, its rapid reoxidation is crucial for the prevention of the catalyst loss; this is of critical importance for the highly corrosive conditions of high sulfide concentration [44, 47]. Instability of the undoped $Mn_xO_y$ under high sulfide concentration was confirmed by the ICP-OES analysis of the supporting electrolyte, which detected an increase in total dissolved manganese concentration at the end of the experiment (i.e., $0.52 \pm 0.01$ mg $L^{-1}$). Thus, although the Ti/TiO$_2$ NTA $Mn_xO_y$ anode was completely stable in the conditions relevant for sewage treatment (i.e., circumneutral pH, 2 mM HS$^-$) investigated in our previous study [22], highly concentrated sulfide solutions caused a rapid and complete dissolution of the active coating, even under anodic polarization. The concentration of the released $Mn^{2+}$ was drastically decreased to $0.09 \pm 0.01$ mg $L^{-1}$ after the Mo doping, and to $0.04 \pm 0.01$ mg $L^{-1}$ for the $Mn_xO_y$-Mo-PANI coating. It should be noted that these tests were conducted in an extremely corrosive environment of 200 mM of sulfide (6.8 g $L^{-1}$ HS$^-$). Moreover, PANI doping slightly improved the sulfide removal rates, from $0.35\pm0.01$ $h^{-1}$ for the $Mn_xO_y$ coating to $0.45\pm0.01$ $h^{-1}$ for the $Mn_xO_y$-Mo-PANI coating, due to the higher surface area caused by its porous morphology. Finally, the improved charge transfer of the coating helps to reduce the energy requirements of the system. Even though the $E_{eo}$ of the system equipped with the Ti/TiO$_2$ NTA $Mn_xO_y$-Mo electrodes was already decreased compared to the pure $Mn_xO_y$, i.e., 3.49 Wh $m^{-3}$ for Ti/TiO$_2$ NTA $Mn_xO_y$ vs only 1.1 Wh m$^{-3}$ for Ti/TiO$_2$ NTA Mn$_x$O$_y$-Mo, it was further reduced once PANI was added to the coating to 0.82 Wh m$^{-3}$ for Ti/TiO$_2$ NTA Mn$_x$O$_y$-Mo-PANI anode.

### 3.3. Impact of operating parameters of sulfide oxidation at the Ti/TiO$_2$ NTA Mn$_x$O$_y$-Mo-PANI anode

#### 3.3.1. Role of dissolved molecular oxygen

Sulfide-laden waste streams are not always anaerobic. Often, they contain molecular oxygen, which may affect the electrocatalytic oxidation of sulfide at Ti/TiO$_2$ NTA Mn$_x$O$_y$-Mo-PANI anodes. First, dissolved oxygen can oxidize hydrogen sulfide into dissolved sulfur species, thus reducing the selectivity towards the production of elemental sulfur [48]. Second, oxygen in the electrolyte can get adsorbed into the oxygen vacancies in the Mn$_x$O$_y$ coating and enhance its catalytic activity through the formation of reactive surface oxygen [49, 50]. To determine the role of the dissolved oxygen in sulfide oxidation, electrocatalytic sulfide removal was performed in anaerobic conditions or under gentle oxygen purge.

As can be seen from **Figure 4a**, sulfide concentration in the solution purged with oxygen remained unchanged in the absence of the anode, meaning that sulfide oxidation with the molecular oxygen is characterized by extremely slow kinetics. Sulfide oxidation only occurred when the Ti/TiO$_2$ NTA Mn$_x$O$_y$-Mo-PANI anode was employed, reaching ∼ 40% removal efficiency under both aerobic and anaerobic conditions. Indeed, the oxidation of sulfide with the manganese oxide is characterized by significantly faster kinetics when compared to its oxidation with the molecular oxygen [48, 51]. Therefore, the presence of oxygen had no effect on the catalytic activity of the Mn$_x$O$_y$-Mo-PANI coating, and sulfide oxidation occurred due to its interaction with active manganese oxide sites.

*3.3.2. Impact of anode potential*

The impact of the anode potential on sulfide oxidation was determined by performing the experiments in the OC and at 1.4 V/SHE, 1.6 V/SHE and 1.8 V/SHE (**Figure 4b**). The tested potential range was increased compared to the previous study (i.e., 0.4 V/SHE – 0.8 V/SHE) [22] in an attempt to enhance the catalyst regeneration and ensure a complete stability of the coating at high sulfide concentration. While anodic potentials above 1.2 V/SHE cause the oxidation of the $Mn_xO_y$ and its subsequent release as permanganate ion (i.e., $MnO_4^-$) [52], the presence of high sulfide concentration in the electrolyte imposes reducing conditions, which prevent the overoxidation and loss of the catalyst. Indeed, no permanganate could be detected even at the highest applied potential (i.e., 1.8 V/SHE).

Without the anodic polarization, sulfide concentration rapidly dropped in the OC by 10 % within 30 min and then remained constant until the end of the experiment (**Figure 4b**). This trend indicates complete dissolution of the $Mn_xO_y$ coating, which occurred as a result of its reduction by the sulfide ion according to the following reaction [44]:

$$Mn^{II} \rightarrow Mn^{2+}_{aq} + \text{new surface site} \qquad (eq.\ 6)$$

This reaction proceeds until the complete depletion of the $Mn_xO_y$ coating. The dissolution of the coating was further confirmed through the analysis of the supporting electrolyte using ICP-OES. The results revealed a significant increase in the total dissolved manganese concentration by the end of the experiment (i.e., 1.3 mg $L^{-1}$ of total dissolved manganese). In our previous work, we demonstrated that electron transfer between the redox couples of Mn and Mo (i.e., Mn III/Mn IV and Mo IV/Mo VI) can stabilize the coating in the OC, preventing its reductive dissolution [23]. Despite the stabilizing effect of the Mo doping on the $Mn_xO_y$ coating, strong affinity of the $Mn_xO_y$ towards the oxidation of sulfide, coupled with the high concentration of sulfide (i.e., 20 mM HS$^-$), ultimately led to complete dissolution of the catalyst in the OC.

The polarization of the Ti/TiO$_2$ NTA Mn$_x$O$_y$-Mo-PANI at 1.4 V/SHE proved to be effective in achieving complete stabilization of the coating, thereby preventing catalyst loss. This successful stabilization was accomplished by applying the potential to restore the reduced Mn II to its initial oxidation state. Application of anodic potential enabled steady sulfide removal, reaching 35% removal efficiency after 2 h of the experiment (**Figure 4b**). Further increase in potential to 1.6 V/SHE and 1.8 V/SHE had a limited impact on the sulfide oxidation kinetics and efficiency. Hence, increase of the potential above 1.4 V/SHE is not needed, as it increases the energy requirements of the system from 3.04 W h m$^{-3}$ at 1.2 V/SHE to 5.6 W h m$^{-3}$ at 1.4 V/SHE, and 9.4 W h m$^{-3}$ at 1.6 V/SHE, without significantly improving the sulfide oxidation kinetics.

*3.3.3. Impact of supporting electrolyte*

The impact of anions, typically present in water or in biogas scrubbing solution, was investigated in this study. Experiments were conducted using NaOH, Na$_3$PO$_4$, Na$_2$CO$_3$, and NaCl electrolytes with matching ionic strength and pH (20 mM, pH ~12). Among these electrolytes, the slowest sulfide oxidation kinetics were observed in the phosphate-based electrolyte (**Figure 4c**). The negative effect of PO$_4^{3-}$ on sulfide oxidation can be attributed to the adsorption ability of the manganese oxides towards phosphate [53]. When phosphate ions are adsorbed onto the manganese oxide surface, they block the active sites of the catalyst, thereby reducing its overall activity [54]. Additionally, anodic polarization of the Mn$_x$O$_y$ coating can facilitate the uptake of phosphate anions through electrosorption, exacerbating the decline of its activity [55].

Although the kinetics of sulfide oxidation were somewhat improved in the NaOH compared to Na$_3$PO$_4$ solution (i.e., from 0.17 h$^{-1}$ for 20 mM NaOH to 0.33 h$^{-1}$ for 20 mM Na$_3$PO$_4$), the process was still significantly slower than in the NaCl or $Na_2CO_3$ supporting electrolytes (**Figure 4c**). Lower oxidation rates could be attributed to the corrosion of the $Mn_xO_y$ caused by the extremely basic pH of the NaOH solution. Several studies demonstrated that $OH^-$ can accelerate the reductive dissolution of the $Mn_xO_y$ by binding to the Mn III sites and leading to the formation of soluble $Mn(OH)_6^{3-}$ [56-58]. Since 0.1 M – 1 M NaOH solution can be used for biogas desulfurization, electrocatalytic sulfide oxidation with $Mn_xO_y$-based anodes will exhibit a comparatively reduced reaction rate under these conditions [59].

Carbonate and chloride-based supporting electrolytes yielded comparable sulfide oxidation kinetics and removal efficiencies (**Figure 4c**). Considering the affinity of free chlorine towards sulfide oxidation, this implies that the $Ti/TiO_2$ NTA $Mn_xO_y$-Mo-PANI anode suppressed chlorine evolution even at potential as high as 1.4 V/SHE [60]. Indeed, free chlorine concentration measured in the absence of sulfide at this potential was negligible, thus excluding the possibility of the formation of toxic chlorinated by-products; formation of chlorinated by-products expected at commercial anodes is particularly problematic for the treatment of industrial sulfide-laden waste streams. On the other hand, although the presence of $CO_3^{2-}$ could potentially passivate the $Mn_xO_y$ catalyst through the precipitation of the carbonate minerals (i.e., $MnCO_3$) at the anode surface, similar performance of the $Ti/TiO_2$ NTA $Mn_xO_y$-Mo-PANI anode in $Na_2CO_3$ and NaCl supporting electrolytes demonstrates that $CO_3^{2-}$ ion did not have any passivating effect on the $Mn_xO_y$ catalyst activity [61]. Biogas upgrading, which targets the simultaneous removal of $H_2S$ and $CO_2$, is often achieved through potassium carbonate scrubbing in which biogas is exposed to 0.1 - 0.6 M $CO_3^{2-}$ containing solution to yield high purity biomethane [62]. Hence, excellent performance of the $Ti/TiO_2$ NTA $Mn_xO_y$-Mo-PANI anode in carbonate-based electrolyte suggests high feasibility of applying the developed anode for electrochemical removal of sulfide from real scrubbing solutions.

*3.3.4. Impact of ionic strength*

The impact of ionic strength of the supporting electrolyte on the sulfide oxidation kinetics was determined by performing the sulfide oxidation experiments in 20 mM, 100 mM, 500 mM $Na_2CO_3$ (**Figure 4d**). $Na_2CO_3$ concentration had no effect on the sulfide oxidation, which further confirms the absence of any passivating effect imposed by the $CO_3^{2-}$ ion. Besides that, it also proves the purely catalytical nature of sulfide removal. If sulfide oxidation was relying on the faradaic charge transfer, supporting electrolyte with higher conductivity would have accelerated the removal process [63]. Finally, it is worth noting that the ionic strength of the electrolyte did not have a significant impact on sulfide oxidation. This observation suggests that the catalytic reaction between the sulfide ion and the $Mn_xO_y$ catalyst occurs through the formation of an inner sphere complex, which aligns with the mechanism suggested in our previous work [22]. This mechanism of complex formation is not hindered by mass transfer limitations. Instead, the rate of the reaction is primarily governed by the ability of $HS^-$ ions to form complexes with the $Mn_xO_y$ catalyst [62].

*3.3.5. Impact of initial sulfide concentration*

Sulfide oxidation experiments were also performed at different initial sulfide concentrations (i.e., 2 mM, 20 mM and 200 mM $HS^-$) (**Figure 4e**). Sulfide oxidation kinetics and oxidation products at the initial concentrations of 2 mM and 20 mM were identical, which further confirms that sulfide oxidation at the $Ti/TiO_2$ NTA $Mn_xO_y$-Mo-PANI anode is not mass transfer limited. Increase in the initial sulfide concentration to 200 mM accelerated the process kinetics. Higher sulfide oxidation rates could be attributed to the formation of polysulfides (i.e., $S_n^{2-}$), which occurred due to the interaction between the electrodeposited elemental sulfur and sulfide ion [64-66]. During this reaction, sulfide ion performs a nucleophilic attack, leading to the dissolution of the elemental sulfur and its release in the form of polysulfides according to the following reaction [66]:

$$HS^- + \frac{n-1}{8}S_8 \leftrightarrow S_n^{2-} + H^+ \tag{eq. 7}$$

High sulfide concentration (i.e., 200 mM HS$^-$) favored sulfur dissolution and subsequent formation of $S_n^{2-}$. Polysulfides are unstable species that are rapidly converted to other dissolved sulfur species, which explains the increased concentration of thiosulfate (i.e., $S_2O_3^{2-}$) at 200 mM HS$^-$ [65]. The Mn$_x$O$_y$-Mo-PANI coating remained stable even at the highest initial concentration of sulfide, which highlights the importance of Mo and PANI doping.

*3.3. Prevention of electrode passivation and sulfur recovery*

Repeated application of the Ti/TiO$_2$ NTA Mn$_x$O$_y$-Mo-PANI electrode in three consecutive cycles at 1.4 V/SHE was performed to investigate the process of anode passivation and to assess the efficiency of chemical and physical regeneration strategies. Consecutive application of Ti/TiO$_2$ NTA Mn$_x$O$_y$-Mo-PANI anode without any regeneration is characterized by the loss of performance and decrease in sulfide removal rate in each subsequent cycle, from 1.13±0.05 h$^{-1}$ in the first cycle, to 0.56±0.2 h$^{-1}$ in the second cycle and 0.08±0.008 h$^{-1}$ in the third cycle (**Figure 5a**). The main product of sulfide oxidation in the first cycle was sulfur deposited at the anode surface. However, in the second cycle sulfur yield was reduced from 85.4% to 33.9%, while polyS fraction increased from < 5% to 49.4% (**Figure 5b**). The $S_n^{2-}$ formation was the major mechanism of sulfide removal in the second cycle. Although this reaction led to the dissolution of the deposited elemental sulfur, sulfide oxidation kinetics decreased even further in the third application cycle. This implies that sulfur dissolution through polysulfide formation is incomplete, and only the outer layer of the deposited sulfur is available for the reaction with HS$^-$.

Physical cleaning of the Ti/TiO$_2$ NTA Mn$_x$O$_y$-Mo-PANI anode in between the applications did not prevent the decline in sulfide oxidation rate constants in each cycle, which were lowered from 0.87±0.09 h$^{-1}$ in the first cycle, to 0.49±0.2 h$^{-1}$ in the second cycle and 0.07±0.01 h$^{-1}$ in the third cycle.

As can be seen from the FESEM image (**Figure S6b**), even after removing the visible sulfur deposit from the anode, the anode surface was still completely covered with a thin layer of sulfur crystals. Although the physical cleaning did not sustain the Ti/TiO$_2$ NTA Mn$_x$O$_y$-Mo-PANI anode performance in terms sulfide removal, it helped to maintain the selectivity of the process towards the production of elemental sulfur, as the generation of polysulfides was suppressed compared with the repeated application of the anode without any cleaning (33.9% of polyS in the second cycle without cleaning vs. 12.1% of polyS in the second cycle with the physical cleaning). The XRD pattern of the particles that were collected after the physical cleaning demonstrated that it was pure sulfur (**Figure S7**). Thus, the recovered sulfur particles represent a valuable industrial product that can be used in various industrial processes. For instance, pure sulfur is an extremely valuable product that used in the production of sulfuric acid [67].

Chemical extraction of sulfur deposited at the Ti/TiO$_2$ NTA Mn$_x$O$_y$-Mo-PANI anode surface enabled robust sulfide oxidation in the consecutive cycles, with sulfide removal rate constants of 1.12±0.32 h$^{-1}$ in the first cycle, 1.46±0.2 h$^{-1}$ in the second cycle and 1.22±0.05 h$^{-1}$ in the third cycle. Cleaning with toluene left the anode surface completely free from any sulfur deposit without compromising the catalytic activity of the Mn$_x$O$_y$-Mo-PANI coating (**Figure S6d**). Furthermore, the generation of polysulfides was avoided and the main final product of sulfide oxidation was elemental sulfur deposited at the anode surface. Once the organic solvent (i.e., toluene) is saturated, sulfur can be easily recrystallized and recovered, enabling the regeneration and continuous reuse of the cleaning solution.

In our previous study, we demonstrated that neutral pH enables continuous sulfide oxidation, while higher pH (i.e., pH 12) hinders the desorption of sulfur particles from the surface, thereby complicating the sulfur recovery process and increasing the likelihood of anode passivation [22].

Hence, the implementation of the suggested electrocatalytic desulfurization approach inevitably leads to anode passivation due to the prevalent basic pH of the biogas scrubbing solutions and industrial sulfide-laden waste streams. Physical cleaning or sulfur dissolution in the form of polyS was shown to be ineffective in preventing passivation, as these methods fail to completely remove the accumulated sulfur. Therefore, continuous sulfide oxidation at the Ti/TiO$_2$ NTA Mn$_x$O$_y$-Mo-PANI anode at basic pH requires periodic cleaning with organic solvent (e.g., toluene), which can be reused by allowing the recrystallization and recovery of sulfur.

## Conclusion

In this study, we synthesized for the first time a PANI and Mo co-doped Ti/TiO$_2$ NTA Mn$_x$O$_y$ anode aiming to investigate the influence of the dopants on electrode stability and activity in the presence of high sulfide concentrations. The incorporation of both Mo and PANI enhanced the catalytic activity of the Mn$_x$O$_y$ coating for sulfide oxidation. These improvements can be attributed to two key factors. Firstly, the inclusion of Mo and PANI led to an increase in the electroactive surface area of the anode. This larger surface area facilitated more active sites for the catalytic reaction, thereby promoting higher sulfide oxidation rates. Secondly, the presence of Mo and PANI resulted in a reduction of the charge transfer resistance. This decrease in resistance enabled more efficient electron transfer during the oxidation process, further enhancing the overall catalytic activity. Moreover, the enhanced conductivity of the coating played a crucial role in the robust regeneration of reduced Mn sites following their reaction with sulfide ions. This conductivity prevented the reductive dissolution of the anode coating, even in the presence of high HS$^-$ concentrations of up to 200 mM.

The impact of various operating parameters (i.e., dissolved oxygen, anodic potential, supporting electrolyte, ionic strength, initial sulfide concentration) on the sulfide (electro)oxidation process was also determined. Dissolved oxygen and ionic strength had limited impact on sulfide oxidation; for example, given that sulfide oxidation occurs through the formation of an inner sphere complex, it was not sensitive to the crowding of the diffusion layer and ionic strength of the solution. Furthermore, an anode potential of 1.2 V/SHE was sufficient to ensure complete electrode stability, and sulfide (electro)oxidation to elemental sulfur ($S^0$) at the energy consumption of only 3.04 Wh $m^{-3}$. Whereas some anions such as phosphate and carbonate were detrimental to the anode performance and stability, carbonate and chloride-based supporting electrolytes resulted in nearly identical sulfide removal rates. Finally, increase in sulfide concentration accelerates the reaction through the formation of polysulfides, however, it may compromise the selectivity of the treatment towards the generation of elemental sulfur.

Application of the Ti/TiO$_2$ NTA Mn$_x$O$_y$-Mo-PANI anode requires chemical dissolution of the deposited sulfur with toluene, from which it can be easily recrystallized and recovered, enabling multiple cycles of solvent reuse. The opportunity for sulfur recovery within the framework of the circular economy can further mitigate the additional cost of chemical (i.e., toluene) usage, especially considering the growing demand for sulfuric acid, driven by a rapid population growth and exponential increase in the need for fertilizers. Given the demonstrated efficiency of the Ti/TiO$_2$ NTA Mn$_x$O$_y$-Mo-PANI anode for sulfide oxidation to elemental sulfur, its (electro)chemical stability even in highly concentrated sulfide solutions, low cost and avoidance of critical raw materials in its production, the developed anode may be an attractive approach for the energy-efficient desulfurization of sulfide-laden waste streams.

## Acknowledgments

The authors would like to acknowledge ERC Starting Grant project ELECTRON4WATER (Three-dimensional nanoelectrochemical systems based on low-cost reduced graphene oxide: the next generation of water treatment systems), project number 714177. ICRA researchers thank funding from CERCA program. The ICS5000 Dionex received support from the CERCA Institute through the CERCAGINYS program, funded by the Spanish Ministry of Science and Innovation.

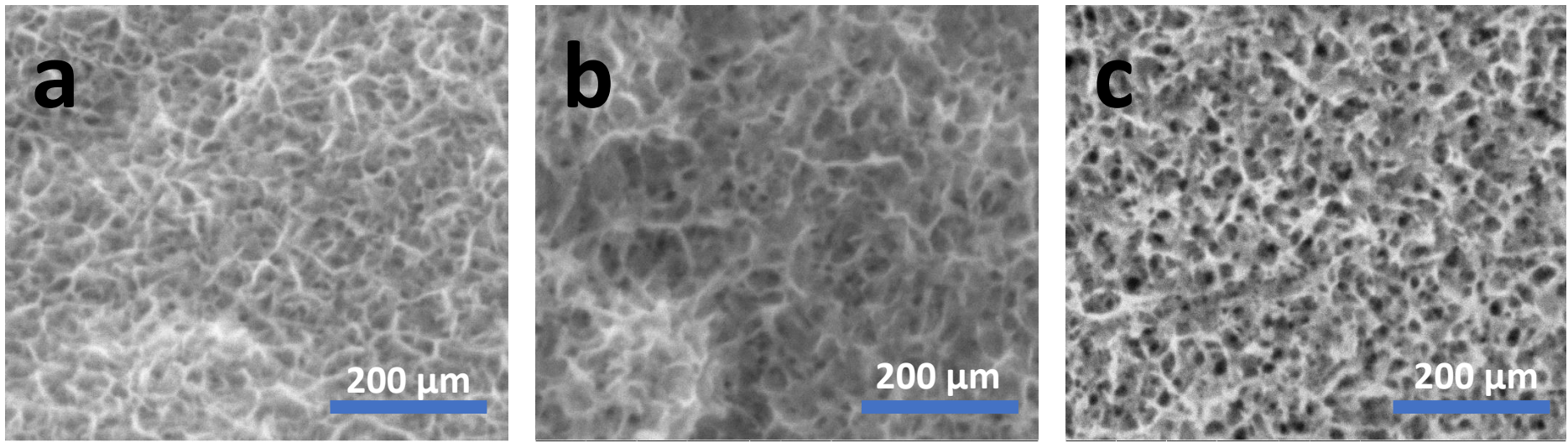

**Figure 1**. FESEM images of the **a)** Ti/TiO$_2$ NTA Mn$_x$O$_y$, **b)** Ti/TiO$_2$ NTA Mn$_x$O$_y$-Mo, and **c)** Ti/TiO$_2$ NTA Mn$_x$O$_y$-Mo-PANI electrode.

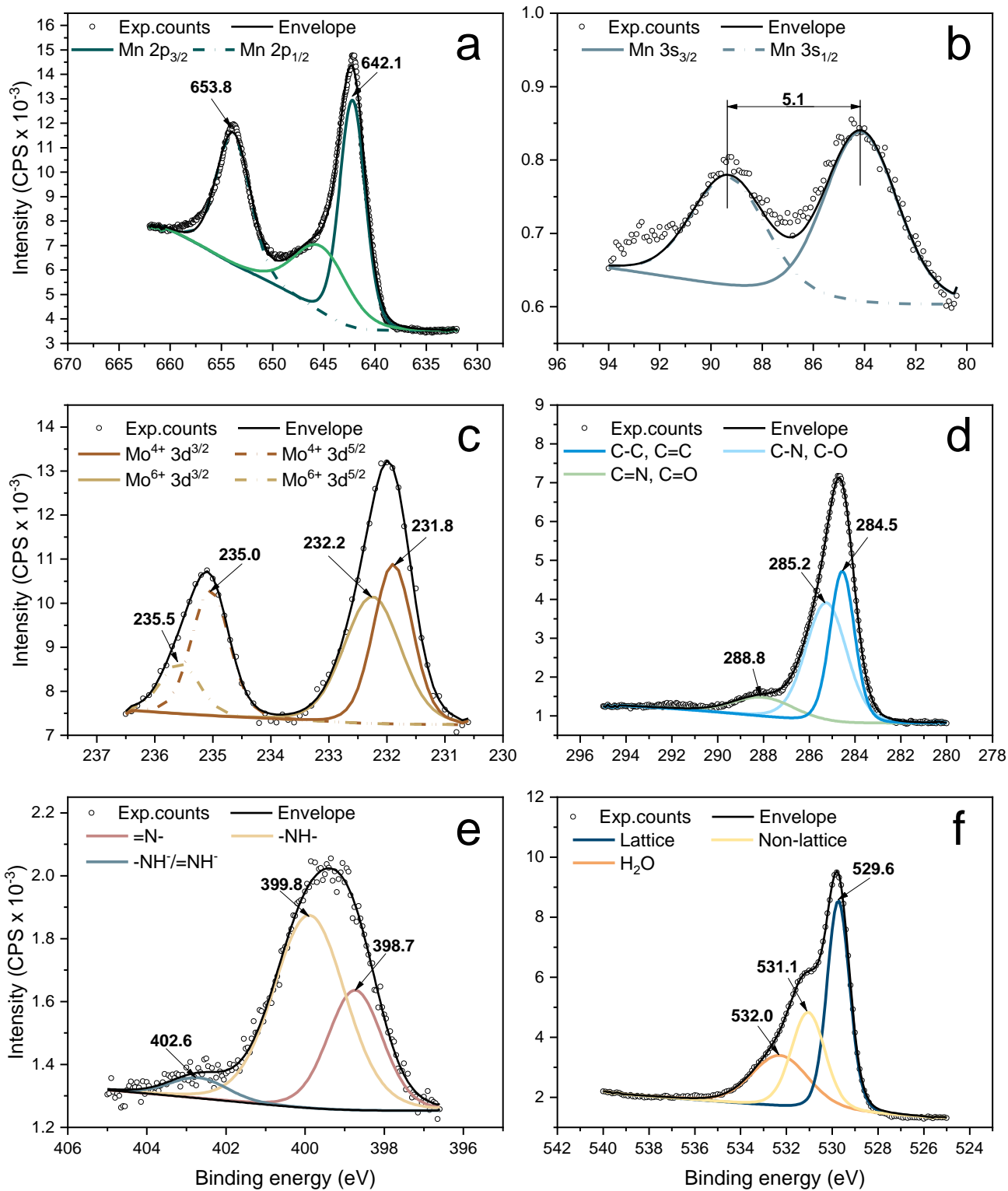

**Figure 2.** XPS spectra displaying the **a)** Mn 2p binding energy, **b)** Mn 3s binding energy, **c)** Mo 3d binding energy, **d)** C 1s binding energy, **e)** N 1s binding energy and **f)** O 1s binding energy of the Ti/TiO$_2$ NTA Mn$_x$O$_y$-Mo-PANI.

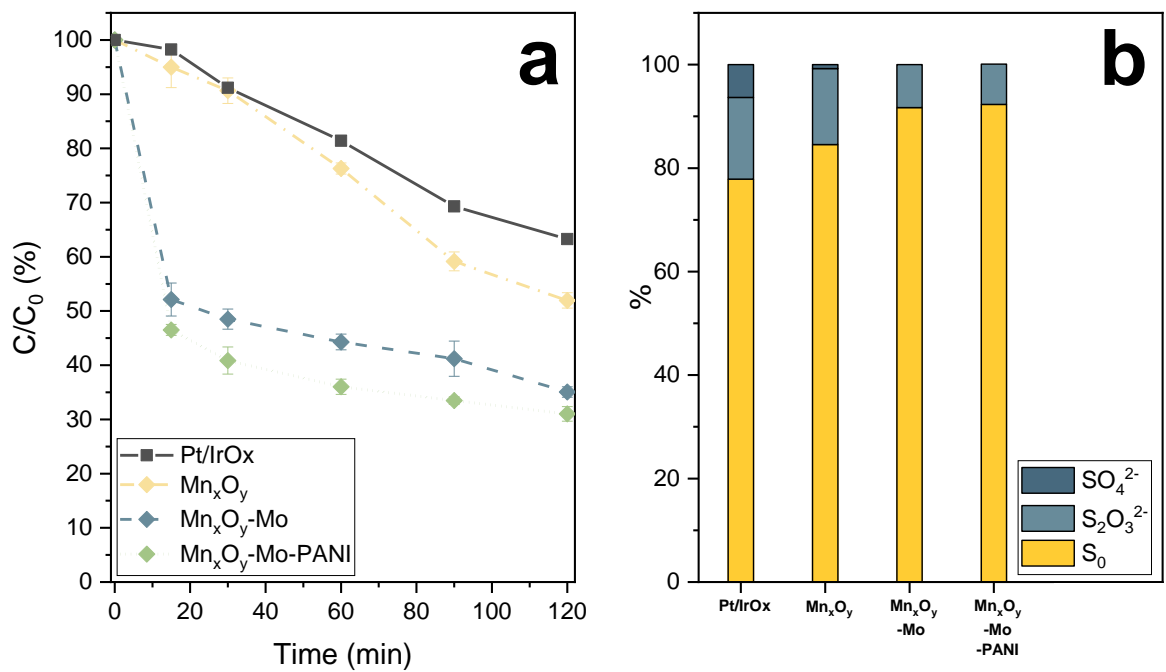

**Figure 3**. **a)** Decrease in HS⁻ concentration (C) normalized to the initial value ($C_0$) during sulfide removal experiment performed in the electrolyte containing 20 mM HS⁻ and 100 mM $NaCO_3$ at 1.4 V/SHE applied to Pt/IrOx electrode, Ti/TiO$_2$NTA electrode coated with $Mn_xO_y$, $Mn_xO_y$-Mo and $Mn_xO_y$-Mo-PANI, and **b)** sulfur species distribution after the (electro)catalytic oxidation of sulfide using different anode materials.

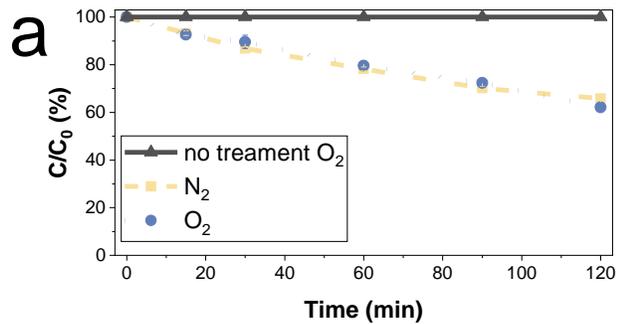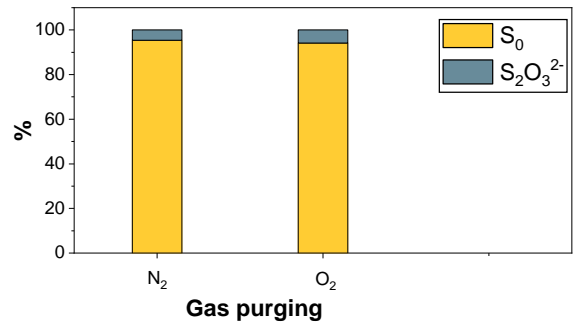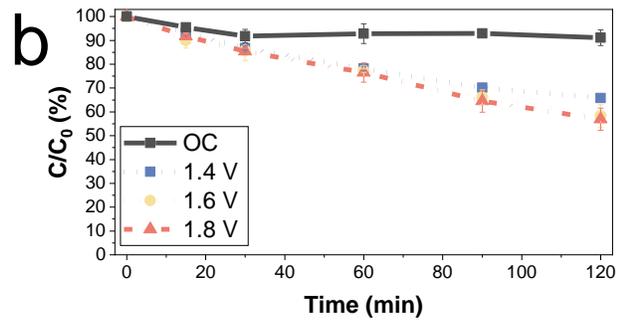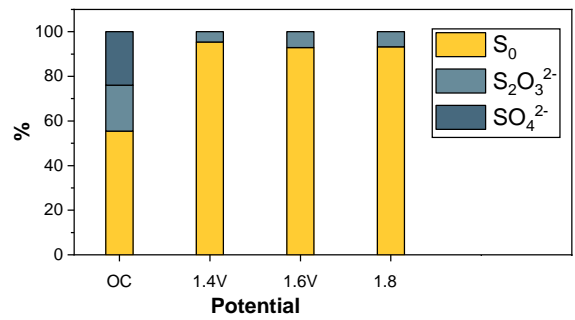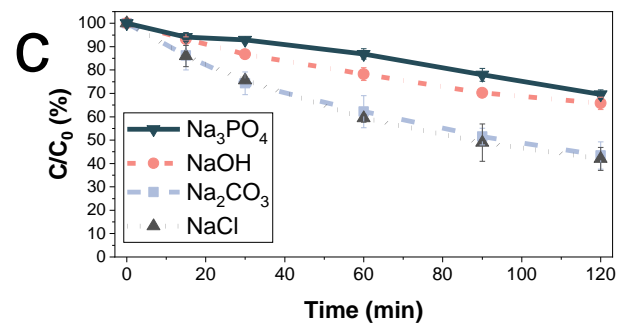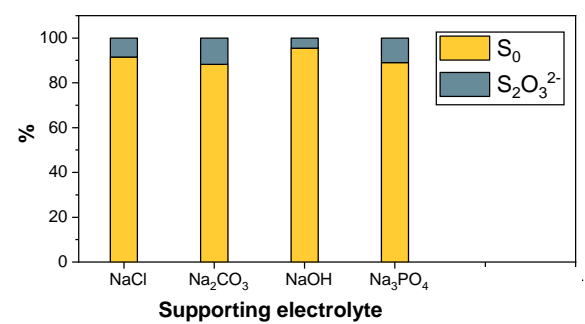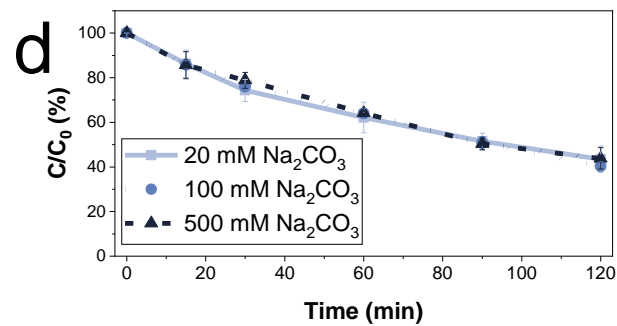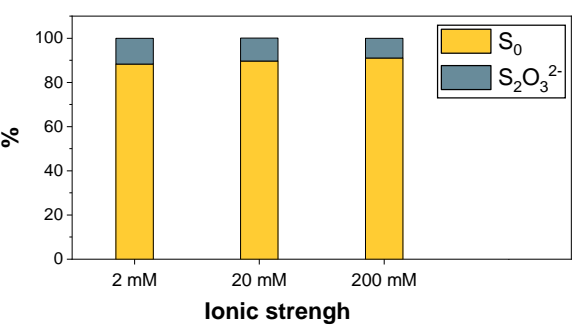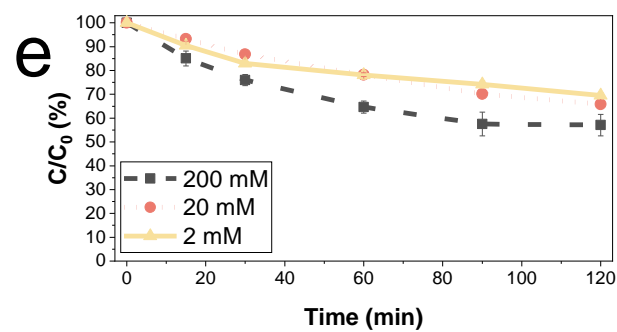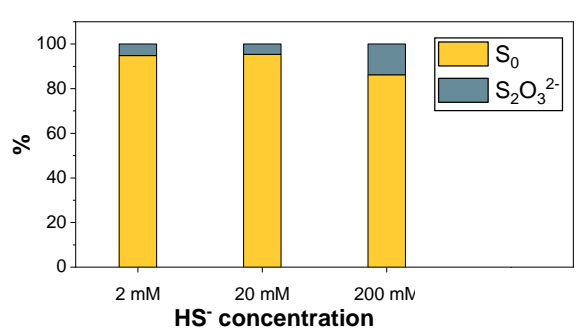

**Figure 4**. Decrease in HS⁻ concentration (C) normalized to the initial value ($C_0$) and sulfur species distribution in the sulfide removal experiment performed **a)** in deoxygenated electrolyte and oxygenated electrolyte with and without the presence of Ti/TiO$_2$ NTA Mn$_x$O$_y$-Mo-PANI anode; **b)** in the open circuit (OC) and electrochemical oxidation experiments performed at various potentials; **c)** in various supporting electrolytes; **d)** in supporting electrolytes of different ionic strength; **e)** at different initial sulfide concentrations.

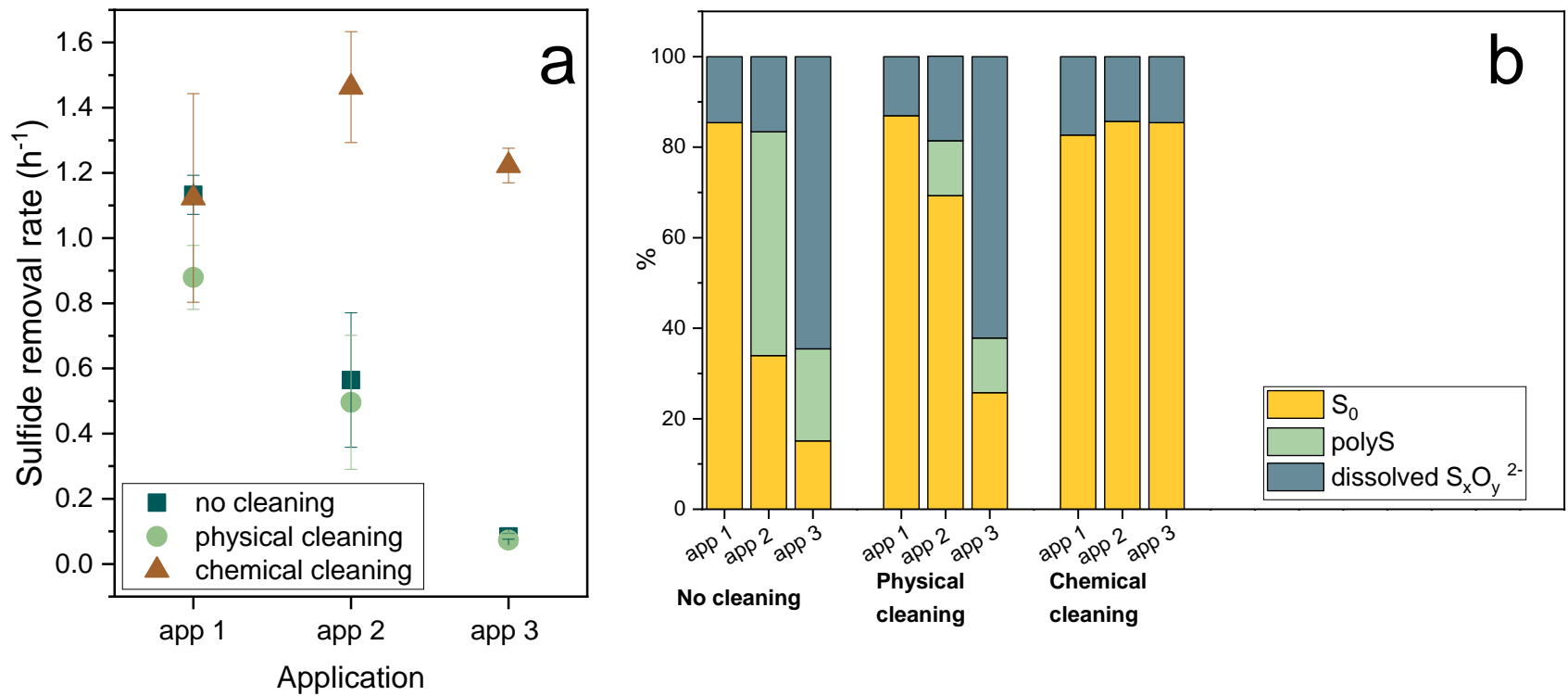

**Figure 5**. **a)** First-order sulfide removal rates (h$^{-1}$) and **b)** sulfur species distribution observed in three subsequent electrochemical oxidation experiments at the Ti/TiO$_2$ NTA Mn$_x$O$_y$-Mo-PANI anode without any cleaning, with physical cleaning and with the chemical sulfur extraction with toluene in between the experiments.

# Supplementary material

**(Electro)catalytic oxidation of sulfide and recovery of elemental sulfur from sulfide-laden streams**

*Natalia Sergienko[a,b], Elizabeth Cuervo Lumbaque[a,b], Jelena Radjenovic [a,c\*]*

*[a]Catalan Institute for Water Research (ICRA-CERCA), Emili Grahit 101, 17003 Girona, Spain*

*[b] University of Girona, Girona, Spain*

*[c]Catalan Institution for Research and Advanced Studies (ICREA), Passeig Lluís Companys 23, 08010 Barcelona, Spain*

*\* Corresponding author:*

*Jelena Radjenovic, Catalan Institute for Water Research (ICRA), c/Emili Grahit, 101, 17003 Girona, Spain*

Phone: + 34 972 18 33 80; Fax: +34 972 18 32 48; E-mail: jradjenovic@icra.cat

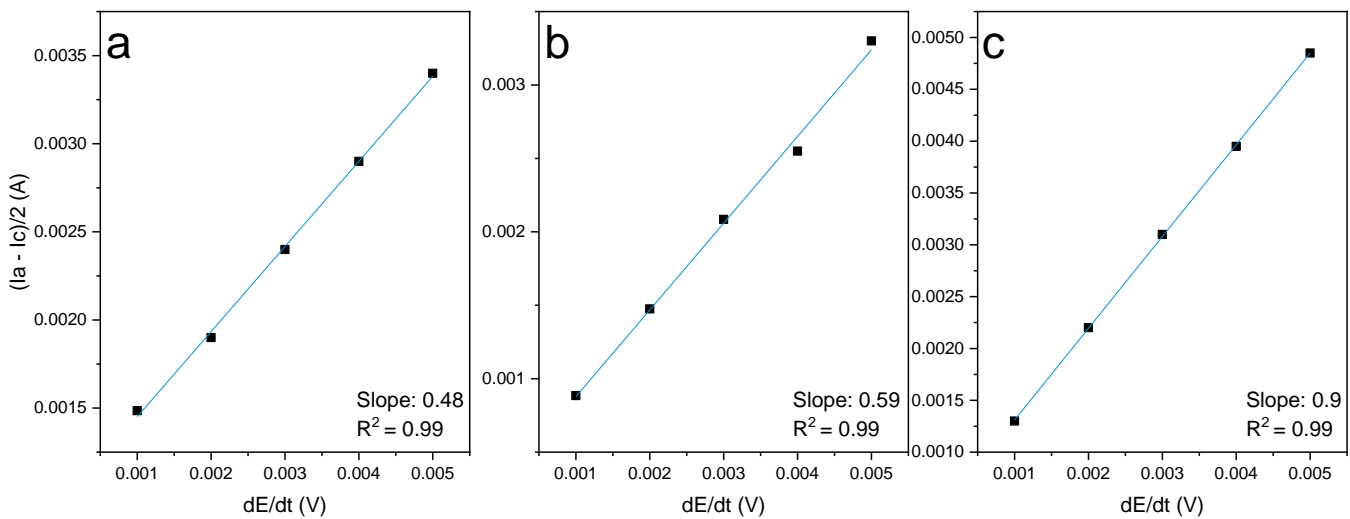

**Figure S1.** Plots of charging currents vs scan rates obtained for **a)** Ti/TiO$_2$ NTA Mn$_x$O$_y$, **b)** Ti/TiO$_2$ NTA Mn$_x$O$_y$-Mo, and **c)** Ti/TiO$_2$ NTA Mn$_x$O$_y$-Mo-PANI.

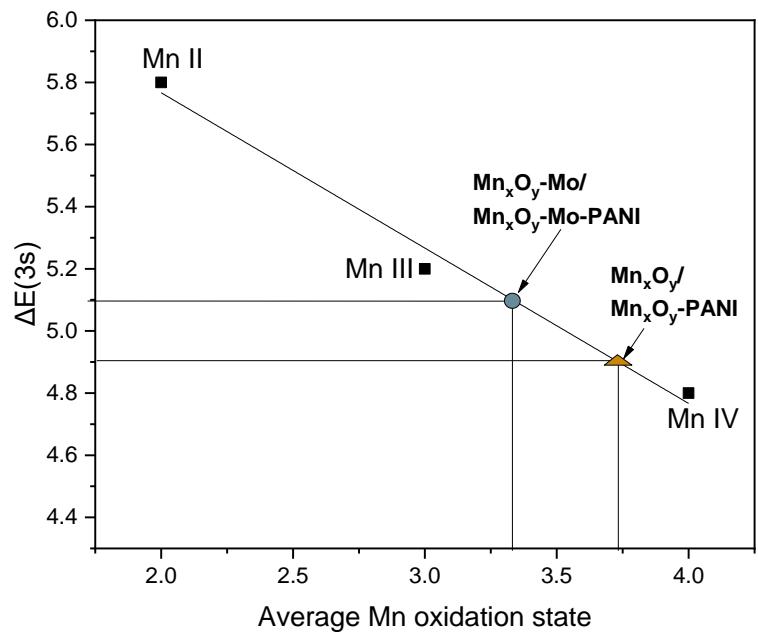

**Figure S2**. XPS linear calibration for quantification of the major Mn oxidation state of $Mn_xO_y$ and $Mn_xO_y$-PANI (brown triangle); $Mn_xO_y$-Mo and $Mn_xO_y$-Mo-PANI (blue circle) based on the $\Delta E(3s)$ binding energy of reference compounds Mn II, Mn III, and Mn IV.

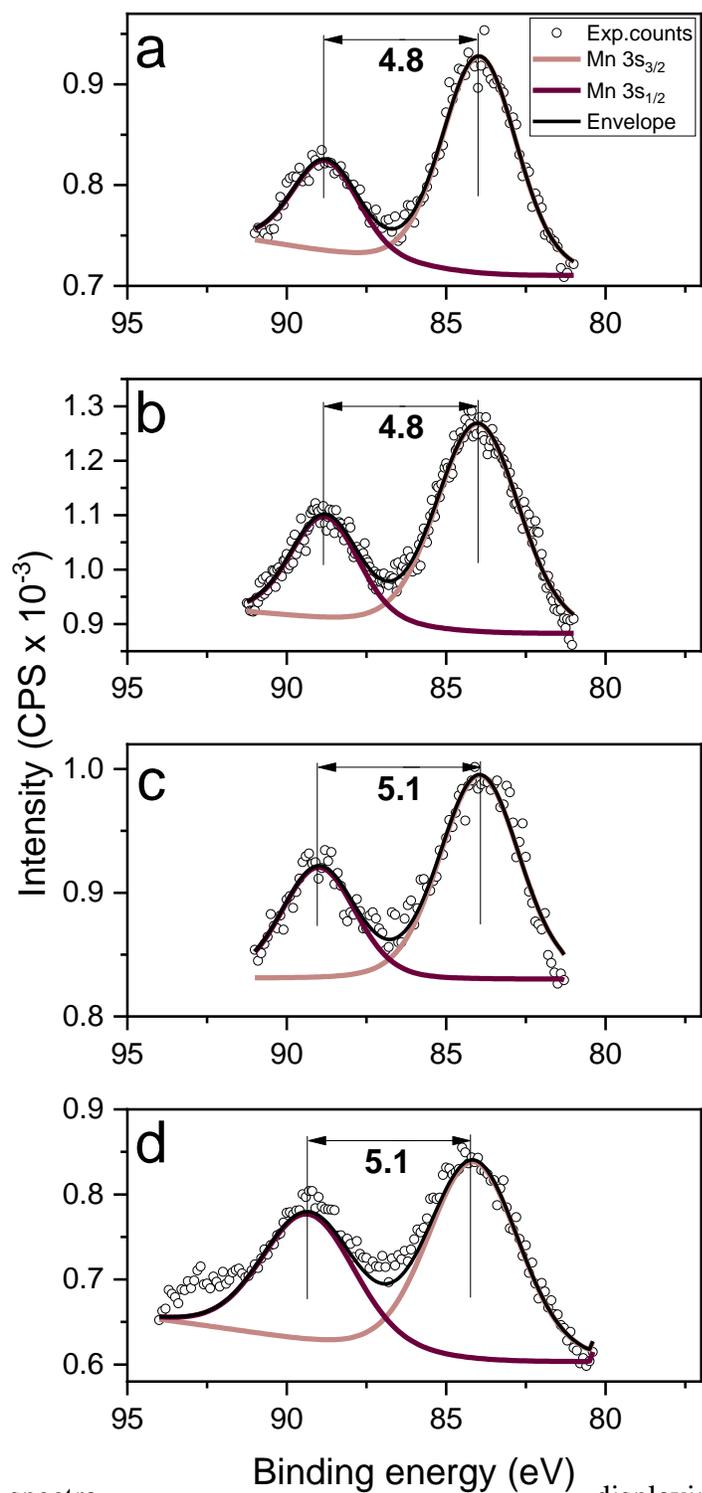

**Figure S3.** XPS spectra displaying the Mn 3s binding energy of **a)** Ti/TiO$_2$ NTA Mn$_x$O$_y$, **b)** Ti/TiO$_2$ NTA Mn$_x$O$_y$-PANI, **c)** Ti/TiO$_2$ NTA Mn$_x$O$_y$-Mo, and **d)** Ti/TiO$_2$ NTA Mn$_x$O$_y$-Mo-PANI.

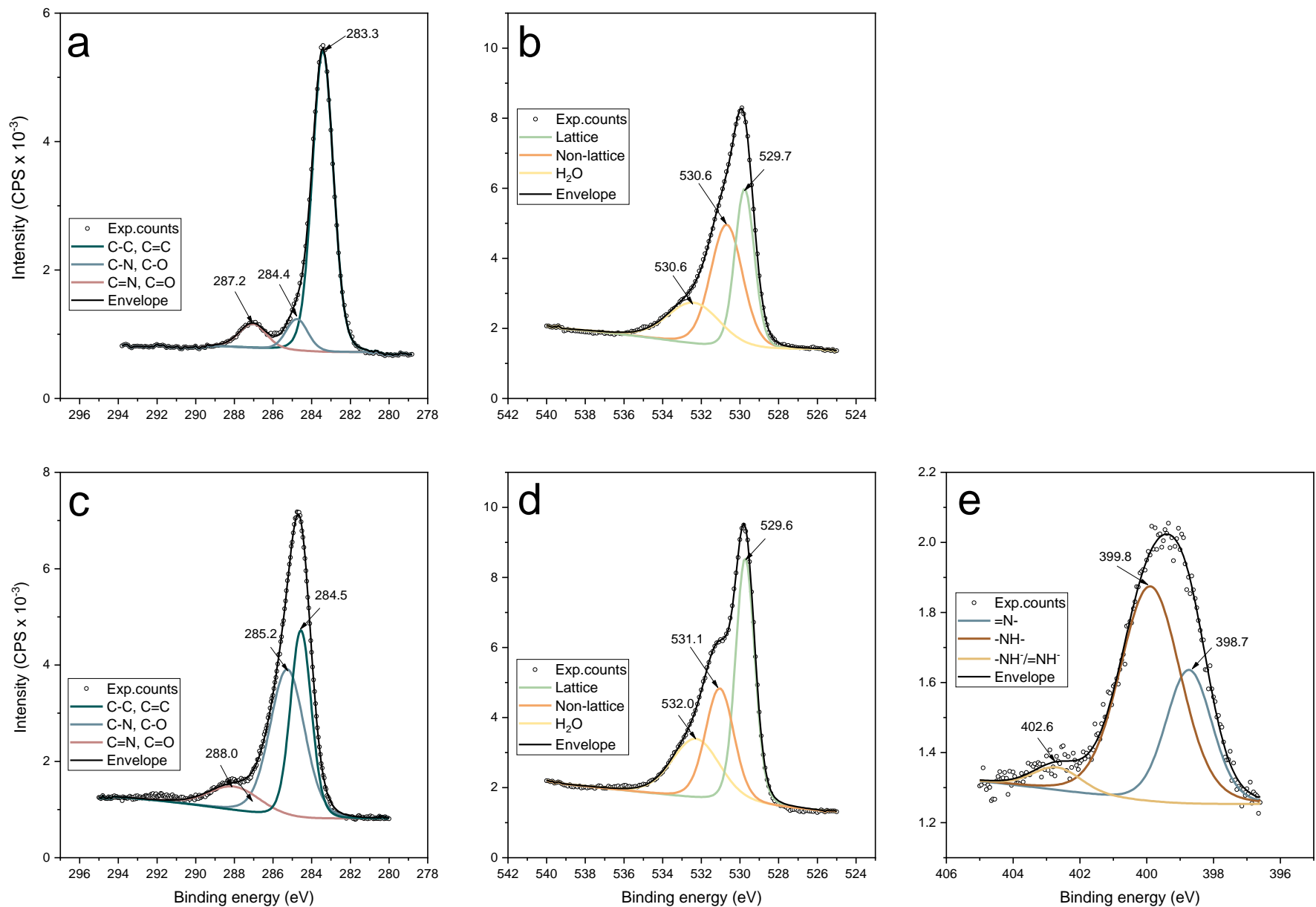

**Figure S4**. XPS spectra displaying **a)** C1s and **b)** O1s binding energy of Ti/TiO$_2$ NTA Mn$_x$O$_y$-Mo, and **c)** C1s, **d)** O1s **e)** N1s of Ti/TiO$_2$ NTA Mn$_x$O$_v$-Mo-PANI.

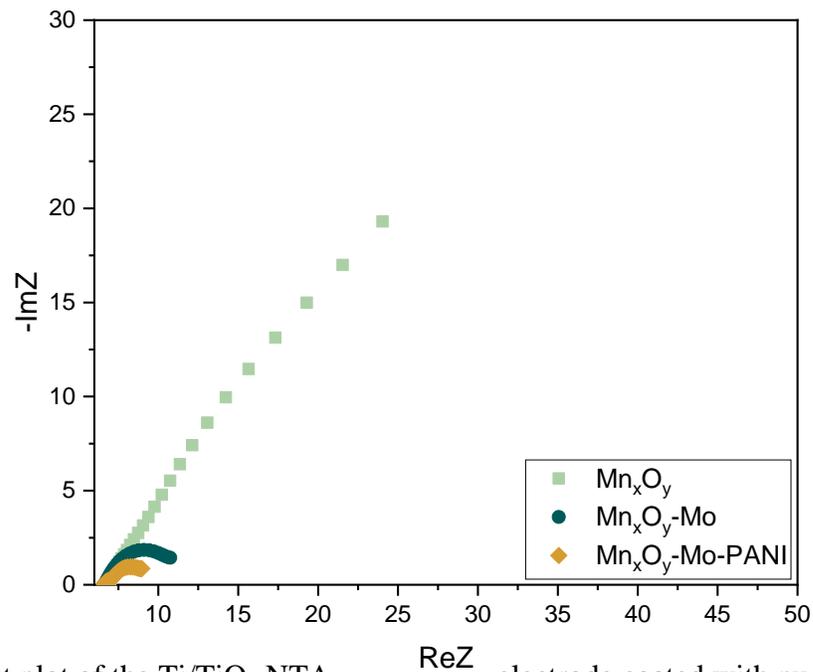

**Figure S5**. Nyquist plot of the Ti/TiO$_2$ NTA electrode coated with pure Mn$_x$O$_y$, Mo-doped Mn$_x$O$_y$ and Mo and PANI co-doped Mn$_x$O$_y$ material in 100 mM NaNO$_3$ at 1.4 V/SHE applied potential in the frequency range of 0.1 Hz−50 kHz.

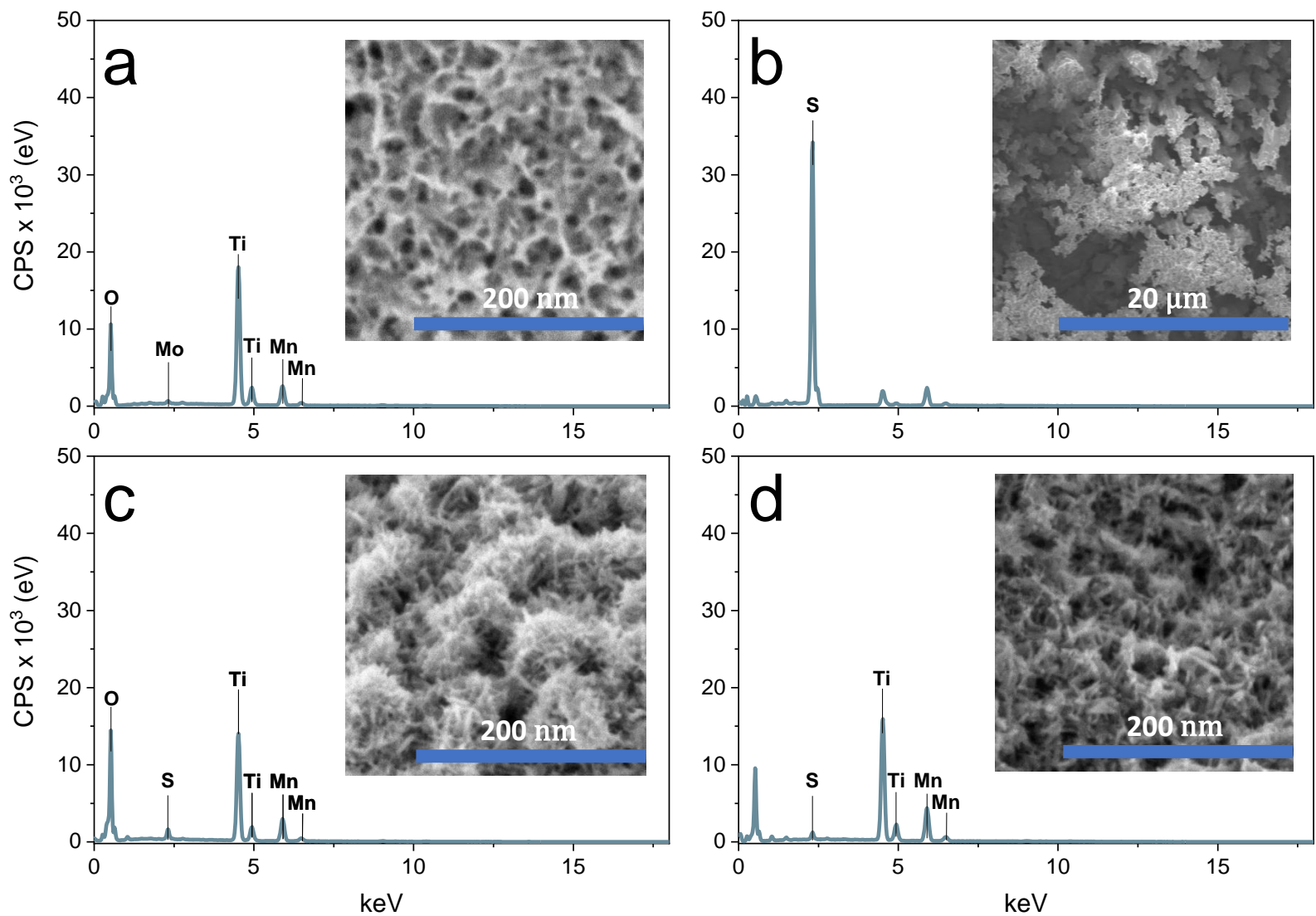

**Figure S6**. FESEM images of **a)** pristine Ti/TiO$_2$ NTA Mn$_x$O$_y$-Mo-PANI, **b)** Ti/TiO$_2$ NTA Mn$_x$O$_y$-Mo-PANI anode after the experiment, with the deposited elemental sulfur, **c)** Ti/TiO$_2$ NTA Mn$_x$O$_y$-Mo-PANI anode after the physical removal of elemental sulfur, **d)** Ti/TiO$_2$ NTA Mn$_x$O$_y$-Mo-PANI anode after the chemical removal of elemental sulfur.

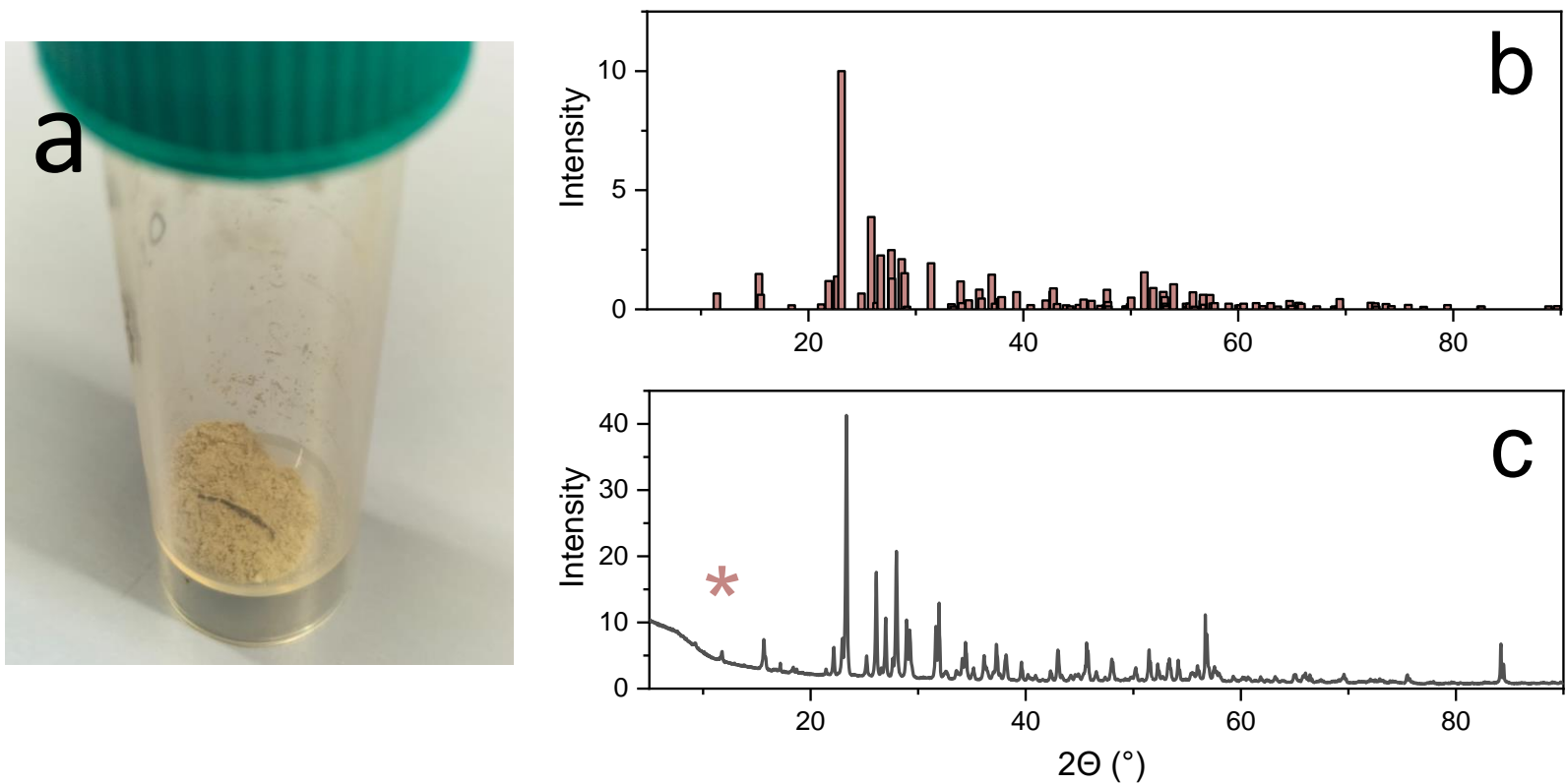

**Figure S7. a)** Images of sulfur particles collected during physical electrode cleaning, **b)** reference XRD pattern of elemental sulfur, and **c)** XRD pattern of collected sulfur particles.

**Table S1.** Binding Energy (eV), FWHM (eV), peak area (counts) and atomic percentages (%) of C1s, O1s and N1s XPS spectra obtained for the Ti/TiO$_2$ NTA Mn$_x$O$_y$-Mo and Ti/TiO$_2$ NTA Mn$_x$O$_y$-Mo-PANI electrodes.

| Element | State | Binding energy, eV | | FWHM, eV | | Peak area, counts | | Concentration, % | |
|---|---|---|---|---|---|---|---|---|---|
| | | Mn$_x$O$_y$-Mo | Mn$_x$O$_y$-Mo-PANI | Mn$_x$O$_y$-Mo | Mn$_x$O$_y$-Mo-PANI | Mn$_x$O$_y$-Mo | Mn$_x$O$_y$-Mo-PANI | Mn$_x$O$_y$-Mo | Mn$_x$O$_y$-Mo-PANI |
| C1s | C-C, C=C | 283.4 | 284.5 | 1.27 | 1.34 | 6652 | 7392 | 82.4 | 52.0 |
| | C-O, C-N | 284.7 | 285.2 | 1.32 | 2.91 | 720.5 | 5564 | 8.93 | 39.1 |
| | C=O, C=N | 287.0 | 288.0 | 1.66 | 2.04 | 696.3 | 1261 | 8.62 | 8.86 |
| O1s | Lattice | 529.7 | 529.6 | 1.19 | 1.16 | 6025 | 6868 | 33.9 | 39.8 |
| | Non-lattice | 530.7 | 531.1 | 1.95 | 1.87 | 7594 | 5405 | 42.7 | 31.3 |
| | H$_2$O | 532.3 | 532.0 | 3.20 | 3.71 | 4175 | 4991 | 23.4 | 28.9 |
| N1s | =N- | - | 398.7 | - | 1.62 | - | 687.9 | - | 30.5 |
| | -NH- | - | 399.8 | - | 2.12 | - | 1447 | - | 64.1 |
| | -NH·, =NH· | - | 402.6 | - | 1.70 | - | 123.7 | - | 5.47 |